\documentclass[authoryear,final,5p,times,twocolumn]{elsarticle}

\usepackage{graphicx}
\usepackage{amsmath,amssymb, amsthm}

\allowdisplaybreaks[3]

\journal{Proceedings of the Royal Society B}

\begin{document}
\begin{frontmatter}

\title{A new perspective on the dynamics of fragmented populations}

\author{Anders Eriksson\corref{cor1}}
\ead{anders.eriksson@marecol.gu.se}
\address{Department of Marine Ecology, University of Gothenburg, SE-43005, Sweden.}
\cortext[cor1]{Corresponding author} 

\begin{abstract}
Understanding the time evolution of fragmented animal populations and their habitats, connected by migration, is a problem of both theoretical and practical interest. This paper presents a method for calculating the time evolution of the habitats' population size distribution from a general stochastic dynamic within each habitat, using a deterministic approximation which becomes exact for an infinite number of habitats. Fragmented populations are usually thought to be characterized by a separation of time scale between, on the one hand, colonization and extinction of habitats and, on the other hand, the local population dynamics within each habitat. The analysis in this paper suggests an alternative view: the effective population dynamic stems from a law of large numbers, where stochastic fluctuations in population size of single habitats are buffered through the dispersal pool so that the global population dynamic remains approximately smooth. For illustration, the deterministic approximation is compared to simulations of a stochastic model with density dependent local recruitment and mortality. The article is concluded with a discussion of the general implications of the results, and possible extensions of the method.
\end{abstract}


\end{frontmatter}

%****************************************************************************************************************
\section{Introduction}
%****************************************************************************************************************

The habitats of many animal populations are geographically divided into numerous small patches, either because of human activities or because the animals natural habitats are patchy. Understanding the dynamics of such populations is a problem of great theoretical and practical interest. Isolated small habitats are often extinction prone, because of, e.g., inbreeding in combination with demographic and environmental stochasticity \citep{hanski99}. When the habitats are connected into a network by migration, local populations may still be extinction prone, but the whole population can be sustained because empty habitats are recolonized by migrants from surrounding occupied habitats. When the migration rate is high, also the local populations are stabilized by an inflow of immigrants (the rescue effect).

There are two main approaches to modeling such populations: First, simplified models that may be treated analytically, and, second, more detailed models where simulations are necessary. In the framework of \citet{levin74}, the full population dynamic is simplified by treating each patch as either occupied or empty. The rates of patch colonization and extinction are expressed as functions of the total fraction of occupied patches in the population, or as functions of the number of neighbouring patches in spatially explicit models. This reduction is generally motivated by a separation of time scale between colonization and extinction of habitats and the population dynamics within the habitats \citep{levin74,hanski_gyllenberg93,lande_etal98}. Many of the mechanisms that can be observed in natural populations (e.g. Allee and rescue effects) have correspondences in Levin's framework, and the qualitative behaviour of such models is well understood \citep[see, e.g.,][]{hanski00,Etienne:2000,harding_mcnamara02,zhou_wang04,Zhou:2004,Taylor:2005,Martcheva:2007}. 
More detailed models incorporate e.g. the effect of correlations between neighbouring patches, leading to non-uniform patterns of colonized patches \citep{hui_li04,Roy:2008}, and of demographic stochasticity within the local habitats \citep{Keeling:2002}. The study of such models generally requires computer simulations \citep[but see][for an exception]{hui_li04}.

Models that describe the metapopulation dynamic as a function of the fraction of occupied patches have no explicit connection to the population dynamic within each patch, i.e. it is not apparent how individual births, deaths, or migration events translate into effective colonization and extinction rates. Starting with \citet{hanski_gyllenberg93}, several authors have attempted to bridge the gap between detailed local population dynamic and the dynamic at the overall population level \citep[e.g.][]{Drechsler:1997,lande_etal98,Nachman:2000,Lopez:2001}. This paper contributes to these efforts by deriving, from first principles, how the global stochastic dynamic leads to an effective deterministic time evolution of the habitats' population size distribution. This gives additional insight in the dynamics of fragmented populations. The only critical assumption in this derivation is that the number of habitats is large enough. Especially, the method can be used for any local stochastic population dynamic.

The rest of the paper is organized as follows: In section~\ref{sec:pop_model} the stochastic population model is presented; in section~\ref{sec:approx_dynamic} the method is developed from the full stochastic dynamic of the population; in section~\ref{sec:example} the method is compared to stochastic simulations of a model with local density dependent recruitment and mortality. Finally, section~\ref{sec:discussion} discusses the general implications of the results, and possible extensions of the method.

%****************************************************************************************************************
\section{Population model}
\label{sec:pop_model}
%****************************************************************************************************************

\begin{figure}[t]
\centering
\includegraphics[width=252pt,clip]{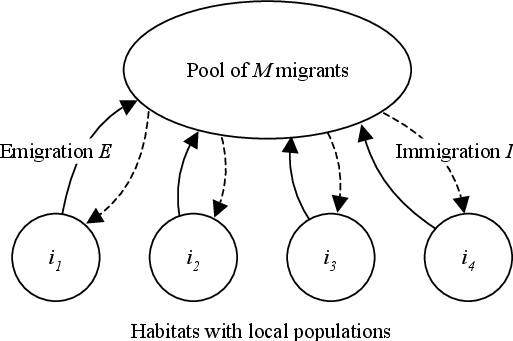}
\caption{\label{fig:pop_illustr}
Illustration of the population dynamic of a patchy population. In each habitat, there are $i_k$ individuals in habitat $k$, and $M$ emigrants in the dispersal pool.
} 
\end{figure}

Consider a population with $N$ habitats (patches), all equal. In each patch, the population changes according to a Poisson process, where the rate of transition from $i$ individuals to $j$ individuals is given by $T_{ji}$. As an important special case, we have the standard birth-death dynamic, where individuals are born and die according to Poisson processes with rates $B_{i}$ and $D_{i}$, respectively, where $i$ is the number of individuals in the patch. In the general model, this corresponds to taking
\begin{align}
	T_{ij} = \begin{cases}
		B_j & \text{when $j = i - 1$} \\
		D_j & \text{when $j = i + 1$,} \\
		   0     & \text{else.}
	\end{cases}
\end{align}
In addition to the local population dynamic, the number of individuals in each habitat can change because some individuals emigrate from the habitat to other patches, or because immigrants arrive from other patches. Following \citet{hanski_gyllenberg93},  the migration process is modeled as follows: Individuals emigrate from a patch at rate $E_i$, where $i$ is the patch population size. If the individuals migrate independently and with constant rate, $E_i$ is proportional to $i$, but more complex migration patterns can be incorporated. If for example individuals move to avoid overcrowding, the migration rate becomes density dependent.
The emigrants enter a common dispersal pool, containing the migrants from all patches that have not yet reached their target patch. Each migrant stays an exponentially distributed time in the pool, with expected value $1/\alpha$, before reaching the target habitat, which is chosen with equal probability among all habitats. This process is illustrated in Figure~\ref{fig:pop_illustr}. 

Migration may fail if the individuals die before reaching the new habitat; this is modeled by the rate $\nu$ of dying per unit of time during dispersal. In practice, the probability of successful migration depends on the background mortality of the individuals, on the time it spends in the dispersal pool, and on additional perils that the individuals may be exposed to during dispersal (e.g. increased risk of predation due to lack of cover, etc.). 
Thus, if there are $M$ migrants, $(\alpha + \nu) M$ individuals leave the dispersal pool per unit of time, and the rate of immigration to a given patch is $I = \alpha M/N$.

%----------------------------------------------------------------------------------------------------------------
%----------------------------------------------------------------------------------------------------------------

%****************************************************************************************************************
\section{Approximative dynamic}
\label{sec:approx_dynamic}
%****************************************************************************************************************

Since the local population dynamic is assumed to be the same within all habitats, it is sufficient to count the number of habitats with a given number of individuals rather than keeping track of the number of individuals in each habitat. Let $n_i$ denote the number of habitats with $i$ inhabitants at a given time. Since the total number of patches is constant, we have $\sum_{i=0}^\infty n_i = N$ at all times. When the number of individuals change from $i$ to $j$ in some habitat, $n_i$ decreases with one and $n_j$ increases with one. From the model in section~\ref{sec:pop_model}, we can write down a set of Master equations for the probability $\rho$ of observing counts $n_0,n_1,\dots$ and $M$ individuals in the dispersal pool (see the appendix). In principle, these equations can be integrated analytically. In practice, however, because the number of possible configurations grows exponentially with the number of patches, it is not possible to use the full Master equation except when the number of habitats and the maximum population size are both small.

When the number of habitats is large enough that the total population size is large, 
even though the number of individuals per habitat may be small,
we can use a different approach. In this case, we can expect that the probability $\rho$ of observing a given distribution of habitat population sizes and number of migrants changes only a little when a single habitat (or a habitat and the migration pool) changes its population size. It is important here to emphasize that we make no assumptions on the size of the changes to the population size of any single habitat, only on that there are sufficiently many habitats that they form a statistical ensemble.
Hence, on the level of the population, this leads only to a single increment and decrement in the count of the number of patches with a given number of individuals. For instance, it is perfectly possible to have big jumps in the population number of individual patches in the model without violating the smoothness of the distribution of habitat population size.

It is now convenient to express the frequency of habitat population sizes in terms of the scaled frequency $f_i = n_i/N$ and the patch immigration rate $I = \alpha M/N$. The probability $\tilde\rho$ of observing $f_0, f_1, \ldots$ and $I$ is related to the probability $\rho$ by
\begin{align}
	\tilde{\rho}(f_0, f_1, \ldots, I) = \rho(N f_0, N f_1, \dots, N I/\alpha) .
\end{align}
In contrast to $n_i$ and $M$, the distribution of $f_i$ and $I$ can be expected to become approximately independent of $N$ for a large number of habitats. A change of a single count $n_i$ leads only to the change $1/N$ in $f_i$. Similarly, a single increment of $M$ leads to the change $\alpha/N$ in $I$.
Since for large $N$ the probability $\tilde\rho$ is an approximately continuous function of $f_i$ and $I$, we can expand the full Master equation of $\tilde\rho$ in terms of $N^{-1}$ to obtain a partial differential equation for $\tilde\rho$ (see Appendix).
Taking only the leading term (ignoring terms of order $N^{-1}$ or higher), we obtain a deterministic dynamic for $f_i$ and $I$, given by 
\begin{align}\label{eq:NME1}
	& \partial_t f_i = 
   \sum_{j = 0,j \ne i}^\infty \big[ T_{ij} f_j -  T_{ji} f_i\big] + I(t) f_{i-1} +
   E_{i+1}f_{i+1} - \big[ E_i + I(t) \big] f_i 
   \nonumber\\
	& \partial_t I = - (\alpha + \nu) I + \alpha \sum_{k=0}^\infty E_k f_k .
\end{align}
In the limit of infinitely many patches, Eq.~(\ref{eq:NME1}) gives an exact description of the dynamic. 

When the migrants spend a short time in the dispersal state compared to the normal birth-death dynamic, i.e. when $\alpha$ is large, $I$ rapidly adjusts so that  $\partial_t I$ becomes approximately zero, i.e.
\begin{align}
	I(t) \simeq \frac{\alpha}{\alpha + \nu}\sum_{k=0}^\infty E_k f_k(t)
\end{align}
at each moment in time. Inserting this into Eq.~(\ref{eq:NME1}) yields a nonlinear Master equation for the patch dynamic,
\begin{align}\label{eq:master_mig2}
	\partial_t f_i =& 
   \sum_{j = 0,j \ne i}^\infty \big[ T_{ij} f_j -  T_{ji} f_i\big] +
   E_{i+1}f_{i+1} - E_i  f_i \nonumber\\
   &+\, \frac{\alpha}{\alpha + \nu} \sum_{k=0}^\infty E_k f_k \left(f_{i-1} - f_i\right).
\end{align}
In this equation, $\alpha/(\alpha + \nu)$ is the probability that a migrant successfully reaches a new habitat. Depending on the biological details of the dispersal, which are not explicitly modeled, $\nu$ may depend on $\alpha$.
The stationary solution of Eq.~\ref{eq:master_mig2} is similar to the result by \citet{Nachman:2000}. He does not, however, derive the equations from the full stochastic dynamic.
In the next section, I investigate the accuracy of this approximation by comparing the solution of the non-linear Master equation to explicit stochastic simulations of a population where the local dynamic is governed by simple growth and density dependent mortality.

%****************************************************************************************************************
\section{Simple density dependent dynamics}
\label{sec:example}
%****************************************************************************************************************

As a simple but illustrative example, we consider a population
with constant birth rate and migration rate per capita, and density dependent mortality. The birth, death, and migration rates of a patch with $i$ individuals are
\begin{eqnarray}
B_{i} &=& r i, \nonumber\\
D_{i} &=& \mu i+ (r-\mu)i^{2}/K, \nonumber\\
E_{i} &=& m i, \label{eq:simple_example_rates}
\end{eqnarray}
respectively, where $K$ is the carrying capacity. 
The parameters where chosen to give modest positive growth rate ($r = 1.05$ and $\mu = 1$) at low numbers. We will assume that births and deaths during migration can be ignored ($\nu \approx 0$), so that Eq.~(\ref{eq:master_mig2}) applies. In order to investigate the accuracy of the theory, we will compare stochastic simulations of this dynamic to the solution of Eq.~(\ref{eq:master_mig2}), first with respect to the stationary states, for different $m$ and $K$, and then to the full time evolution.

%================================================================================================================
\subsection{Quasi-stationary states}
%================================================================================================================

%----------------------------------------------------------------------------------------------------------------
\begin{figure}[t]
\centering
\includegraphics[width=252pt]{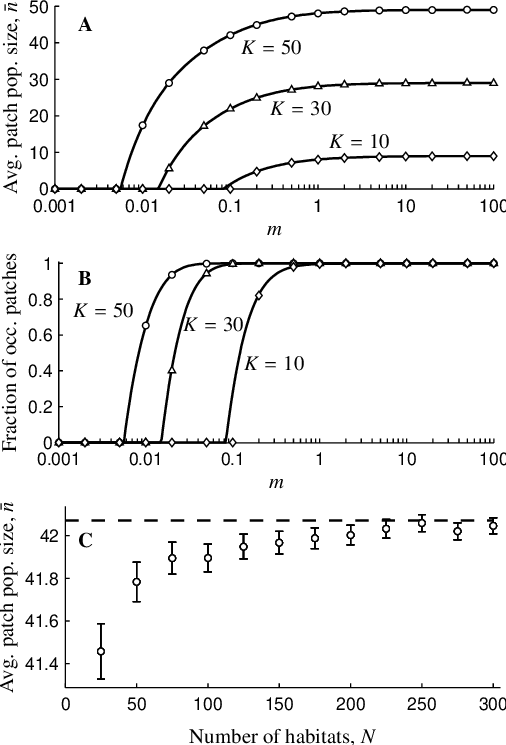}
\caption{\label{fig:n_p_of_m}
A: The average patch population size $\bar n$ at equilibrium,
as a function of migration rate $m$ from theory (solid line), and
from simulations for $K=50$ (circles), $K=30$ (triangles), and $K=10$ (diamonds). The parameters are $N=1,000$, $\mu=1$, and $r=1.05$.
B: The fraction of colonized patches as a function of migration rate $m$ from theory (solid line), and
from simulations (symbols).
C: The average patch population size for an infinite number of habitats (dashed line) compared to simulations of finite number of habitats for $K=50$ (circles, 95\% approximate confidence intervals).
} 
\end{figure}
%----------------------------------------------------------------------------------------------------------------

Fig.~\ref{fig:n_p_of_m} shows the average patch population size, $\bar n$, and the fraction of occupied patches as a function of the migration rate, from explicit simulations of the stochastic dynamic of $N = 1,000$ patches and from numerical solution of Eq.~(\ref{eq:master_mig2}), for three different values of the carrying capacity $K$. The initial state for each simulation point was generated by distributing $NK$ individuals randomly among the patches, so that the initial average number of individuals per patch is $K$. The populations were simulated to time $1,000$, and the mean population size and the fraction occupied patches were calculated as time-averages over the interval $200 \le t \le 1,000$. If the population went extinct before the end of the simulation interval, the mean population size and fraction of occupied patches were taken to be zero.

\citet{hanski_zhang93} analyzed a deterministic version of this model, where the local dynamic is replaced with the deterministic counterpart. Using the technique of \citet{hanski_gyllenberg93}, they found that the dynamic converges to $\bar{n} = K$ and all patches occupied.
When the migration rate $m$ is large, the average patch population size and the fraction of occupied patches are close to the values predicted by the deterministic local dynamic, but for small $m$ the differences are significant; especially, with stochastic local dynamic both patch population size and the fraction of occupied patches decrease with decreasing migration rate, and for each value of $K$ there is a critical migration rate below which the population cannot sustain itself and becomes extinct. For all values of $m$, the stationary solution to Eq.~(\ref{eq:master_mig2}) is in excellent agreement with the simulations. This illustrates the importance of accounting for stochasticity in the local dynamic, even though the global population changes deterministically.

Fig.~\ref{fig:n_p_of_m}C shows how the average patch population size typically depends on the number of patches (see caption for parameter values), compared to the theoretical value in a population with an infinite number of patches. The average is increasing with increasing number of patches, but also for small $N$ the relative difference is not large.

%================================================================================================================
\subsection{Transient dynamics}
%================================================================================================================

An import advantage of the theory presented in this paper is that it not only gives the equilibrium points of the stochastic population dynamic, but the complete time evolution of the population. In this section, we compare the time evolution of the full stochastic simulations to the solution of Eq.~(\ref{eq:master_mig2}).

\begin{figure}[t]
\centering
\includegraphics{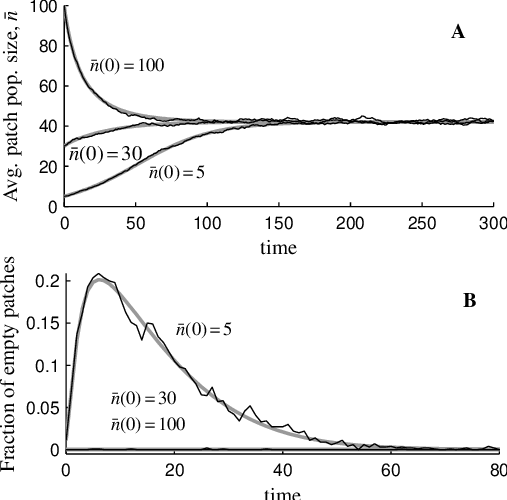}
\caption{\label{fig:avg_n_ex1}
A: Comparison of the average patch population size
in simulation trajectories of $N=10^{3}$ patches (black lines) to
the solution of the nonlinear Master equation, Eq.~(\ref{eq:master_mig2}),
starting from the same initial distribution of patch population sizes
as the simulations (grey lines), for three initial average patch population
sizes ($\bar{n}=5$, $\bar{n}=30$, and $\bar{n}=100$). The parameters
are $\mu=1$, $r=1.05$, $K=50$, and $m=0.1$.
B: Time evolution of the fraction of empty patches, for the same trajectories as panel A. For $\bar{n} = 30$ and $\bar{n} = 100$ the fraction empty patches is approximately zero at all times.
} 
\end{figure}

Fig.~\ref{fig:avg_n_ex1}A shows the time evolution of the average patch population size from simulations of $N = 1,000$ patches (black lines), with migration rate $m = 0.1$, starting from populations with three values of the average patch population size [$\bar{n}(0)=5$, $\bar{n}(0)=30$, and $\bar{n}(0)=100$]. The initial populations were produced by distributing $N\bar{n}(0)$ individuals randomly over the islands. Also shown are the corresponding solutions to the nonlinear Master equation, (Eq.~\ref{eq:master_mig2}, grey lines), starting from the same initial patch population size distribution as the corresponding simulations. Simulations and theory are in close agreement, except some fluctuations in the simulations from the finite number of patches. These fluctuations become more pronounced around the stationary point of the dynamic. This can be expected since away from the equilibrium, relatively strong deterministic forces dominate the dynamic, but close to the equilibrium the dynamic is dominated by the stochastic components.

%----------------------------------------------------------------------------------------------------------------
%----------------------------------------------------------------------------------------------------------------

Panel B of the same figure shows the time evolution of the fraction of empty patches for $n=5$ (for the other values of $\bar{n}$, the fraction of empty patches is approximately zero at all times), from simulations and the nonlinear Master equation. As before, there are small differences from the stochastic fluctuations in the simulation. The number of patches is initially increasing, before reaching a maximum and eventually decreasing towards zero. This is a marked difference from the deterministic theory, that predicts a steady decrease towards zero.
 
\begin{figure}[t]
\centering
\includegraphics{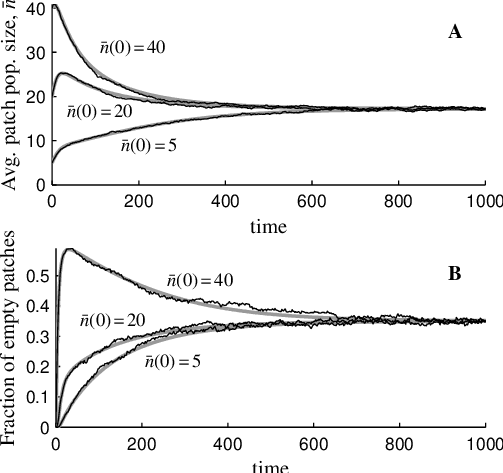}
\caption{\label{fig:m_0.01}
A: Comparison of the average patch population size
in simulation trajectories of $N=10^{3}$ patches (black lines) to
the solution of the nonlinear Master equation, Eq.~(\ref{eq:master_mig2}),
starting from the same initial distribution of patch population sizes
as the simulations (grey lines), for three initial average patch population
sizes ($\bar{n}=5$, $\bar{n}=20$, and $\bar{n}=40$). The parameters
are $\mu=1$, $r=1.05$, $K=50$, and $m=0.01$.
B: The time evolution of the fraction of empty patches. 
} 
\end{figure}

Fig.~\ref{fig:m_0.01} shows the time evolution for a smaller migration rate, $m = 0.01$. This value is close to the critical migration rate for these parameters (c.f. Fig.~\ref{fig:n_p_of_m}). As a consequence, the average patch population size at the stationary point is much lower than for $m = 0.1$ ($\approx 18$ compared to $\approx 42$), and the fluctuations in the population are larger. Therefore, the simulation curves in Fig.~\ref{fig:m_0.01} are averages over ten independent realizations of the stochastic dynamic, starting from the same initial distribution.

%----------------------------------------------------------------------------------------------------------------
%----------------------------------------------------------------------------------------------------------------

\begin{figure}[t]
\centering
\includegraphics{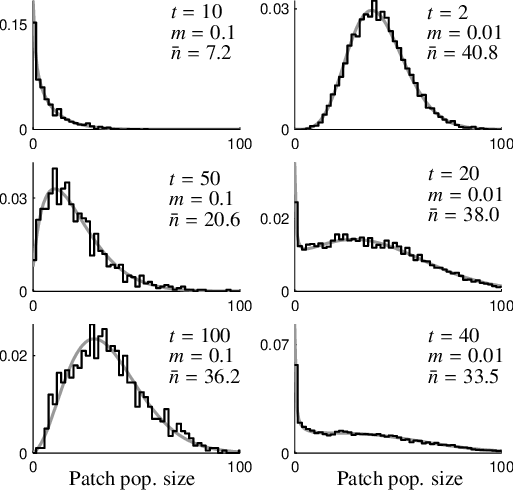}
\caption{\label{fig:distr_n_ex1}
Normalized histogram of patch population sizes from simulations (black lines) compared to the solution of the non-linear Master equation (grey lines). Left column: $m = 0.1$, started from $\bar{n}(0) = 5$, for $t = 10$,  $t = 50$, and  $t = 100$ (from the simulation in Fig.~\ref{fig:avg_n_ex1}). Right column: $m = 0.01$ and $\bar{n}(0) = 40$, for $t = 2$,  $t = 20$, and  $t = 40$ (from the simulation in Fig.~\ref{fig:m_0.01}). For increased legibility, the simulation histograms where binned in groups of two.
} 
\end{figure}

As a final comparison, Fig.~\ref{fig:distr_n_ex1} shows the distribution of patch sizes in the simulations to the solution of the nonlinear Master equation at different times, from the simulations in Fig.~\ref{fig:avg_n_ex1} and Fig.~\ref{fig:m_0.01}. Because of the finite number of patches ($N=10^{3}$) there is significant spread around the prediction of Eq.~(\ref{eq:master_mig2}), but as can be seen in the right column, where the distribution is averaged over ten independent realizations, Eq.~(\ref{eq:master_mig2}) predicts the expected distribution quite well.

%----------------------------------------------------------------------------------------------------------------
%----------------------------------------------------------------------------------------------------------------

%****************************************************************************************************************
\section{Discussion}
\label{sec:discussion}
%****************************************************************************************************************

The model presented in this article makes several assumptions that are idealizations of more realistic scenarios. Most of these are not necessary, but were introduced in order to simplify the description of the method.
For example, it is assumed that all habitats have the same size, availability of food etc., so that the dynamic is the same. In many cases, habitats are not equal but can vary significantly in terms of size and quality \citep{hanski_gyllenberg93,Hanski:1994}. It is straightforward to extend the model presented in this paper to incorporate population structure, where the parameters governing the local patch dynamic may vary between patches.

In addition, it is assumed that individuals migrate independently of others, although the rate of migration can have an arbitrary dependence on the local patch population size. In biologically more realistic situations, several individuals could leave and arrive together. In addition, it may happen that both emigration and immigration leads to (or is caused by) a disruption of the local population, e.g. a conflict \citep{Johst:1997}. It is possible to include such processes directly into the equations of motion by modeling them with a probability transition matrix that connects the states before and after the event.

Local colonization and extinction events are correctly captured by Eq.~(\ref{eq:NME1}), and if the dynamic is such that the population is driven to extinction in a deterministic manner, e.g. because deaths are systematically more frequent than births, these equations will show this as well. The risk of extinction of the whole population from demographic stochasticity cannot be calculated, however, since the derivations assume that the number of migrants in the dispersal pool is large. In order to study such events it is necessary to extend the calculation in the appendix to terms of the order $N^{-1}$. The resulting partial differential equation is a diffusion approximation for the distribution of habitat population sizes, which can be used to approximately calculate the expected time to extinction of the whole population and more accurately estimate average population sizes.

One of the main conceptual difficulties in modeling the population dynamic in terms of population extinction and colonization events is in defining when an empty patch is `colonized': Is it when the first individuals arrives at the path, as in \citep{Keeling:2002}, or should the patch be considered colonized first when the number of individuals is typical of a non-empty patch, as in \citep{lande_etal98}? The method in this paper avoids this issue completely, since it deals directly with the distribution of the number of individuals in the habitats.
This is especially important when habitats are small but many. In this case, there may not be a clear separation between colonized and empty patches, but rather a continuum of patch population sizes (c.f. Fig.~\ref{fig:distr_n_ex1}).

The standard perception of the population dynamic of fragmented populations has been that the local birth and death dynamic is much more rapid than the rate of extinction and recolonization of patches, so that the number of individuals in the occupied patches approaches a value determined by the number of occupied patches between each patch colonization or extinction event.
The derivations presented in this paper show that this view can be slightly misleading;
rather than resulting from a separation of time scale between local dynamic and patch colonization and extinction, the effective population dynamic stems from a law of large numbers, where local fluctuations are buffered through a common migration pool so that the global dynamic remains smooth. Hence, the change in the fraction of occupied patches and in the average number of individuals on occupied patches are coupled through their shared dependence on the rate of immigration.

Finally, since the only critical assumption behind these results is that the number of habitats is large, the theory is especially suited to further elucidating the interplay of local stochastic dynamics and environmental factors in populations undergoing change.

\section*{Acknowledgements}

This work was funded by the Centre for Theoretical Biology, Gothenburg University. I am indebted to Karin Harding for introducing me to the metapopulation concepts and to the work by Hanski and Gyllenberg, and to Andrea Morf and David Kleinhans for proofreading and comments.

\appendix

%****************************************************************************************************************
\section{Derivation of Eq.~(\ref{eq:NME1})}
\label{sec:appendix}
%****************************************************************************************************************

Our starting point is the time evolution of the probability $\rho$ of having $n_i$ habitats with $i$ individuals, for $i = 0,1,2,\ldots$ and $M$ individuals in the dispersal pool, with the total count normalized as $\sum_{i=0}^\infty n_i = N$.
This gives an exact description of the full stochastic population dynamic. Taking into account the local dynamic in each patch, the migration process and the mortality due to migration, we obtain the Master equation
\begin{align}\label{eq:distr_master_eq}
  & \partial_t \rho(n_0, n_1,\ldots, M) = \nu (M+1) \,\rho(\ldots, M+1) - \nu M \rho  \nonumber\\ 
  &    + \sum_{i,j=0}^\infty \left[ T_{ij} (n_j + 1)\, \rho(\ldots,n_i-1,\ldots,n_j+1,\dots) - T_{ji} n_i \rho \right] \nonumber\\
  &   + \sum_{i=0}^\infty E_{i+1} (n_{i+1} + 1) \, \rho(\ldots,n_i - 1, n_{i+1} + 1, \ldots, M-1)  \nonumber\\
  &   + \sum_{i=1}^\infty \alpha (M+1) \frac{n_{i-1} + 1}{N} \, \rho(\ldots, n_{i-1} + 1, n_i-1, \ldots, M+1) \nonumber\\
  &   - \sum_{i=0}^\infty \left(E_i n_i +  \alpha M \frac{n_i}{N} + \nu M \right) \rho  
\end{align}
for these probabilities. For brevity, we show only the variables date differs from the left-hand side in these expressions. We now express the state in terms of the normalized frequencies $f_i = n_i/N$ and the rate of immigration to a given patch, $I = \alpha M/N$. When the number of habitats is large, each event leads only to a small change in the probability $\tilde\rho(f_0, f_1, \ldots, I)$ (each change is of the order $1/N$). Expanding $\tilde\rho$ to order $1/N$, and ignoring terms of order $1/N^2$ and higher, leads to the deterministic equations of motion for $f_i$ and $I$.
\begin{align}
  & \partial_t \tilde\rho = N \sum_{i,j=0}^\infty \left[ T_{ij} \left(f_j + \frac{1}{N}\right)\left(\tilde\rho - \frac{1}{N}\partial_{f_i}\tilde\rho + \frac{1}{N}\partial_{f_j}\tilde\rho\right)  - T_{ji} f_i \tilde\rho \right] \nonumber\\ 
  &+ N \sum_{i=0}^\infty E_{i+1} \left(f_{i+1} + \frac{1}{N}\right) \left(\tilde\rho - \frac{1}{N}\partial_{f_i}\tilde\rho +\frac{1}{N}\partial_{f_{i+1}}  \tilde\rho - \frac{\alpha}{N}\partial_{I}\tilde\rho \right) \nonumber\\    
&+ N \sum_{i=1}^\infty \left(I + \frac{\alpha}{N}\right) \left(f_{i-1} + \frac{1}{N}\right) 
	\left(\tilde\rho + \frac{1}{N}\partial_{f_{i-1}}\tilde\rho -\frac{1}{N}\partial_{f_{i}}\tilde\rho  +\frac{\alpha}{N}\partial_{I}\tilde\rho \right) \nonumber\\
&+ N \frac{\nu}{\alpha} \left(I + \frac{\alpha}{N}\right) \left(\tilde\rho + \frac{\alpha}{N}\partial_{I}\tilde\rho\right) 
-	N \sum_{i=0}^\infty \left( E_i f_i +  f_i I + \frac{\nu}{\alpha} I \right) \tilde\rho.
\end{align}
Collect terms in increasing order of $N^{-1}$, we find that the coefficients of the leading term (proportional to $N$) cancel, and the remaining terms can be written as
\begin{eqnarray}
  \hspace{-0.5in}&& \partial_t \tilde\rho = \sum_{i,j=0}^\infty T_{ij} \left[ \partial_{f_j} (f_j \tilde\rho) - \partial_{f_i} (f_j \tilde\rho)\right]
  + \partial_I \left[  (\alpha + \nu) I \tilde\rho - \alpha \sum_{i=1}^\infty E_i f_i \tilde\rho \right] \nonumber\\ 
  \hspace{-0.5in} &&+ \sum_{i=0}^\infty \partial_{f_i} \left[ \left(E_i + I\right) f_i \tilde\rho - I f_{i-1} \tilde\rho - E_{i+1} f_{i+1} \tilde\rho  \right]
\end{eqnarray}
where we have ignore all terms of order $N^{-1}$ or higher. Hence, the probability obey a transport equation. Rearranging the terms, we obtain Eq.~(\ref{eq:NME1}).

%****************************************************************************************************************
% Bibliography
%****************************************************************************************************************

\bibliographystyle{authordate1}

\end{document}